# ESTIMATING BIOMASS MIGRATION PARAMETERS BY ANALYZING THE SPATIAL BEHAVIOR OF THE FISHING FLEET

*ESTIMACION DE PARAMETROS MIGRATORIOS DE LA BIOMASA A TRAVES DE UN ANALISIS DEL COMPORTAMIENTO ESPACIAL DE LA FLOTA DE PESCA*

**HUGO SALGADO***
Universidad de Talca

**ARIEL SOTO****
Universidad de Concepción

### Abstract

*In this study, a method will be developed and applied for estimating biological migration parameters of the biomass of a fishery resource by means of a decision analysis of the spatial behavior of the fleet. First, a model of discrete selection is estimated, together with patch capture function. This will allow estimating the biomass availability on each patch. In the second regression, values of biomass are used in order to estimate a model of biological migration between patches. This method is proven in the Chilean jack mackerel fishery. This will allow estimating statistically significant migration parameters, identifying migration patterns.*

**Keywords:** *Biological dispersion, industrial fishing, fishing migration, spatial bioeconomics model, stock distribution.*

**JEL Classification:** *Q22.*

* Profesor Asistente, Departamento de Economía, Universidad de Talca, Talca, Chile. E-mail: *hsalgado@utalca.cl*
** Profesor Instructor, Escuela de Administración y Negocios, Universidad de Concepción; Vicente Méndez 595, Chillán, Chile. E-mail: *arsoto@udec.cl*. Teléfono: +56-042-2207550 (Autor corresponsal).

*** The authors would like to thank for the support provided by Chilean National Committee of Science and Technology (CONICYT) from the Chilean Government. This institution has collaborated with the funding of this research by means of the grant FONDECYT 11090233. Additionally, the authors appreciate the comments provided by Jorge Dresdner and Claudio Parés from the Department of Economy at the University of Concepción, which undoubtedly helped improve the quality of this research.

**Resumen**

*En este artículo se desarrolla y aplica una metodología para estimar parámetros biológicos de migración de la biomasa de un recurso pesquero, por medio del análisis de las decisiones del comportamiento espacial de la flota. Primero se estima un modelo de elección discreta, conjuntamente con una función agregada de captura por parche. Esto permite estimar la disponibilidad de biomasa en cada parche. En la segunda etapa se utilizan los valores de la biomasa para estimar un modelo de migración biológica entre parches. La metodología se aplica a una pesquería del jurel en Chile. Esto permite estimar parámetros de migración estadísticamente significativos, identificando patrones de migración.*

Palabras claves: *Dispersión biológica, pesca industrial, migración pesquera, modelo bioeconómico espacial, distribución del stock.*

Clasificación JEL: *Q22.*

## 1. INTRODUCTION

Bioeconomic analysis in fisheries has experienced remarkable development in recent years, especially in the analysis of the spatial behavior of fishermen and fish stock. This is mainly due to the utilization of Marine Protected Areas (MPA) as a tool for fisheries management. Spatial research has been carried out on both the biological and economics aspects of the fishery but also combining the two areas for optimal bioeconomic policy evaluation. Biological studies characterize the fish resource with emphasis on zonal features such as rates of reproduction and growth (e.g. Janmaat, 2005; Kelly *et al.*, 2000; Jones, 2006)[1]. Economics studies characterized the behavior of fishermen in harvesting decisions, such as determinants of the choice of fishing zones and the search process itself, and moreover, information flows among fishermen on the spatial availability of fish resources in the different fishing zones (e.g. Díaz and Salgado, 2006; Zhang, 2010 among other)[2]. A bioeconomic model combines the spatial behavior of the fish biomass and fishermen to fully capture the dynamics of spatial models that leads to optimal policies on management of the fisheries (e.g. Sanchirico and Wilen, 1999; Sanchirico and Wilen, 2001; Smith and Wilen, 2003).

The use of spatial management tools in fisheries, such as the MPA is not trivial due to the uncertainty about the impacts they may present. Different biological and economic effects are generated by this sort of tools in comparison with others, such as Catch Limits or Territorial Use Rights, for instance. First, when the captures in a fishing area are closed, positive effects on the biomass are expected. However, when the economic behavior of the fishermen is incorporated, it is possible to see different effects outside the target area of regulation; however, some of those effects may be negative. This can subtract the possible biological benefits of creating a reservoir. On the other hand, it is possible that migration characteristics of the stock between zones can make possible the fact that the creation of reservoir zones even generates a double payoff. One of these would be biological due to an increment on the biomass, and one economic, due to the recovery of the biomass. Not only within the reserve area, but also outside of it, leading to an increase in the captures and in the economic outcomes of the fishery (Sanchirico and Wilen 2001).

---

[1] See too: Bohnsack 1998; Wallace 1999; Pezzey, Roberts, and Urdal 2000; Murawski *et al.*, 2000; Rowe, 2001; 2002; Halpern, 2003; Russ, Alcala, and Maypa, 2003; Gell and Roberts, 2003; Layton, Haynie, and Huppert, 2003; Grafton, Kompas, and Lindenmayer, 2005; Denny and Babcock, 2004.

[2] See too: Smith and Wilen, 2003; Smith 2005 and Cartigny *et al.* 2008.

Despite the importance of the spatial analysis in fisheries, the empiric works developed to date have been focused on the economic aspects of the analysis because there is neither information nor adequate knowledge on the biological processes -such as migration in this case- that could determine the behavior of the biomass of the fishing resources in a way that could be incorporated into mathematically treatable bioeconomic models. Thus, empiric studies have focused only on decision processes by the fishermen, ignoring the biology of the resource; or they have assumed that the biology of the resource is known, and calibrated values are used for the simulations.

In Chile, Díaz and Salgado (2006) and Cartigny *et al.* (2008) have analyzed the problem of the spatial distribution of the fishing zones between industrial and artisanal sectors, using tools from the MPA literature. In such works, the authors parameterized a bioeconomic model of a marine reservoir with data from the *anchoveta* (Engraulis ringens Jenyns, 1842) fishery in Northern Chile. By using this model, the different distributions of the fishing zone between industrial and artisanal fishermen are simulated by analyzing the long-term evolution of the biomass and both artisanal and industrial captures in the different fishing areas. The authors conclude that a zoning of the fishing area may attain biological and economic objectives as well as income distribution between both sectors. Thus, these authors divide the fishing zone in two patches and assume certain parameters of biological dispersion between these two zones. In order to counteract the scarce information on this issue, the authors simulate their results for different dispersion rates of the biomass amongst patches, concluding that the dispersion rate is a crucial parameter towards achieving biological and economic objectives in the use of the fishing areas zoning as a tool for fisheries management.

On the other hand, Cartigny *et al.* (2008) developed a Bioeconomic model of dynamic optimization in order to analyze the optimal distribution of a fishing zone between artisanal and industrial sectors, taking into account that the areas reserved for artisanal fishing also have a biological importance due to the presence of zones coastal upwelling, which implies higher rates of biomass growth and a process of biomass dispersion from coastal areas to zones where the industrial sector is permitted to catch. Again, the authors conclude that the biological parameters associated with the growth of the resource and the biomass dispersion rates between patches are fundamental to find an internal solution and a stationary balance in the problem of the dynamic optimization. Due to the lack of information about the biomass migration that allows a realistic application of the model, these authors only present simulations by means of standard parameters from the bioeconomic literature for illustrative purposes. The impossibility of finding biological information that allows the application of this theoretical model to a study case has prompted the development of this research.

Some authors have developed the notion of analyzing the stock of biomass as a latent variable in the decision to capture by fishermen, basically through the inclusion of dummies per period that represent aspects which will remain constant among fishermen (Murdock 2006; Timmins and Murdock 2007; Zhang 2010). For instance, Zhang (2010) developed a model for analyzing the decision of capture effort for fishermen in the Gulf of Mexico. The author develops a three-stage econometric procedure that allows identifying the economic model that explains the determination of fishing effort of the individuals and their labor supply.

The most directly related work to this study is that by Smith *et al.,* (2009). In this article, the authors analyze the importance of considering spatial and dynamic processes in the analysis of the renewable resources management, such as fishery resources. The authors apply a model to the estimation of the biological parameters of migration in the reef-fish fishery in the Gulf of Mexico. The article states that *"…a critical component of the spatial-dynamic systems is the dispersion or diffusion mechanisms that link temporary and spatial components of the model…"* (Smith *et al.,* 2009: 108). However, these authors are aware of the practical difficulties of knowing these components, due to the lack of specific information on the migratory biology of the fishery resources. Nevertheless, they point out that it is possible to develop methods that will allow estimating the parameters of the migration models by observing the behavior of fishermen. That is the contribution of this work, in the sense that it is proposed an estimation method of migratory parameters through disaggregated information based solely on observing the behavior of the fleet. In our model, disaggregated data at the level of the behavior of the fishermen are not required; it suffices to know the aggregated fishing effort and the captures per patch and period. To that end, an input function estimation method of demand proposed by Berry (1994) is used. Additionally, estimation of the function of capture by patches, which will enable the direct identification of the biomass level estimated by patch and period in the first estimation stage.

The organization of this article is as follows: In the next section, the theoretical model that allows estimating the parameters of biological migration is presented and discussed, using aggregated data for the fishery on distribution of effort and captures per patch and fishing area. In Section 3 the most important elements of the Chilean mackerel fishery in Central Southern Chile are presented, as well as data used for the estimation. This for contextualize the example. Section 4 presents and discusses the results of the estimation and finally, Section 5 concludes.

## 2. METHODOLOGY

In this section, the bioeconomic model is presented. Its structural estimation allows identifying economic parameters associated to the functions of capture, as well as biological parameters associated to the growth of the resource and carrying capacity of the system. In the second step, parameters of stock migration between different fishing patches are identified.

In the first step, the assumptions of the biological model that describes both growth and migration of the stock between patches are described. Subsequently, the economic model associated to the technology of captures is presented. Finally, the assumptions of the fishermen's behaviors that allow explaining the observed data and the estimation of the structural parameters of the model are also described.

**Biological model**

The biological component of the model is based on the works developed by the economic literature in order to analyze the migration of fishery resources between patches. Among the literature that analyzes these models, Cartigny *et al.,* (2008); Sanchirico and Wilen, (1999); Sanchirico and Wilen, (2001); Smith, (2005); Smith and Wilen, (2003) and Smith *et al.,* (2009) can be mentioned, amongs others.

It is assumed that the fishing zone is divided in $k \times 1$ patches. The dynamics of the stock in the $k$ patch is given by:

$$x_{t+1}^k = x_t^k + f(x_t^k) - H_t^k + MN_t^k \qquad (1)$$

Where, $x_t^k$ is the stock in the patch, $f(x_t^k)$ is the natural growth of the biomass in patch $k$, and $H_t^k$ is the fishing mortality or capture. $MN$ represents the net migration of the patch, defined as the difference between immigration from other patches and migration to other patches.

Patches are different spatial environments that contain subpopulations of a same biomass. As it is common in the literature (Smith *et al.*, 2009), it is assumed that the growth of the biomass is logistical:

$$f\left(x_t^k\right) = r x_t^k \left(1 - \frac{x_t^k}{K^k}\right) \qquad (2)$$

Here, $K$ is the carrying capacity of the patch $k$; and $r$ is the instantaneous growth rate. Using the formulation proposed by Sanchirico and Wilen (2001), it is assumed that the net migration presents the following shape:

$$MN_t^k = d_{kk} x_t^k + \sum_{\substack{h=1 \\ h \neq k}}^{K} d_{hk} x_t^k, \quad k = 1,...,n. \qquad (3)$$

In this case, the growth component is omitted (in comparison with the original one) because this is included explicitly in equation (1). On the one hand, $d_{kk}$ represents the migration rate; therefore it is negative. On the other hand, $d_{hk}$ represents dispersion rates between patches *h* and *k*. These very authors accept that this parameter is "in itself very stylized" ignoring many other aspects, but it is "analytically tractable". So, it is possible to rewrite (1) in a matrix form for the *K* patches as:

$$x_{t+1} = x_t + F(x_t)_t - H_t + Dx_t \qquad (4)$$

Equation includes *k* equations where *k* is a vector dimension (*k*) that represents the biomass in each patch. Additionally, $F(\cdot)$ is a diagonal matrix ($k \times k$) whose elements include natural growth rates $f_k(x_k)$ and $D$ is a ($k \times k$) dimension matrix whose elements show dispersion rates among patches. Finally, $H_t$ shows the vector that includes captures in each patch.

For the case analyzed in this work, it is assumed that the dispersion matrix follows a multidirectional scheme with the constraint that migration occurs only to neighboring patches, where each row adds to one. Thus, for a general case, the dispersion matrix shows the following shape:

$$D = \begin{bmatrix} d_{11} & d_{12} & \cdots & 0 \\ d_{21} & d_{22} & \cdots & 0 \\ \vdots & \vdots & \ddots & \vdots \\ 0 & 0 & \cdots & d_{kk} \end{bmatrix}$$

Some representations for this matrix may be as useful as other, simpler ones, equally allowing the measurement of all dispersion parameters for each patch.

Assuming that the migration occurs only between neighboring patches is not an a-priori pre-requisite for modelling; everything will depend on the level of temporary disaggregation of the database. Namely, if it is a very long period, this assumption will be hardly sustainable. However, for every-day data, as in this case, it is a reliable assumption.

## 2.1. Captures-per-patch function

We assume that the capture per patch function has a Cobb-Douglas form, as presented in equation (5). In this functional form, the variable γ represents a technological coefficient, $E_t^k$ represents the aggregated effort in the patch, and *z* represents other variables that may affect the capture.

According to Berry's methodology (1995), the *z* vector possesses the features observed by the econometrist. It must contain the fishermen's characteristics (vessel dimensions, warehouse capacity, engine horsepower, etc.), and the patches' characteristics (sea surface temperature, sea level, wave height, etc.). The choice of these variables will depend on their feasibility, according to the case.

In addition, α, β, γ and ρ are parameters which are assumed as constant among the different patches.

$$H_t^k = \gamma \left(E_t^k\right)^\alpha \left(x_t^k\right)^\beta \left(z_t^k\right)^\rho \qquad (5)$$

**2.2. Selection per fishing area**

Our model assumes that every time fishing is performed, the fishermen must decide in which patch capture effort is performed; and in which patch it is not (and therefore, no captures will be made). Thus, when fishing, the decision of the fishermen must be represented as a dichotomous decision where the selection options are represented by selecting a patch for fishing or not to go fishing.

In order to make this decision, it is assumed that the fisherman performs an estimation of the capture that could be obtained in each patch if an effort is applied ($E_t^k = 1$). This would be given by:

$$\widehat{H}_t^k = \gamma \left(x_t^k\right)^\beta \left(z_t^k\right)^\rho e^{\varepsilon_t^k} \qquad (6)$$

In the previous equation it is assumed that the existing biomass in each patch is known by the fishermen at the time of deciding where to go fishing. This is based on the fact that fishermen possess experience in the development of their work, and added to that the fact that they can share information on the best fishing zones with other fishermen at the port. Therefore, it is assumed that their decision of sailing is rational, i.e. optimized.

Additionally, it is assumed that the fisherman takes into account the net price per capture unit, which also includes the costs of the expected capture unit in the patch. In order to simplify the subsequent estimation of the model, it is assumed that a constant parameter ($\hat{p}^k$) collects information on the price per net ton, net of costs, of capturing in a specific patch.

A price vector is constructed that will sensitize the fisherman's benefits to the distance traveled from the port to the patch. The vector has the form: $\hat{p}_t^k = \left(\overline{p}_t - \hat{c}_t^k\right)$. Here, $\overline{p}_t$ refers to the price of the resource unloaded in the $t$ period, and $\hat{c}_t^k$ is the operating expense (OE) of the fisherman per ton captured $\left(OE_t^k / H_t^k\right)$. The operational expense considered the displacement time per performance (in nautical miles of each vessel). For this, it is necessary that the researcher know an estimated of this performance, the price of fuel, and the distance from each patch to each port. In this manner, the perspective of the researcher is to design the fisherman's exercise as a consumer who declares their preferences for each patch, in function of the net utilities that each one obtains, evaluating the distance between the benefit per capture and the operating expenses of going fishing to said patch.

Thus, in order to decide in which patch the fishing effort must be applied, the fisherman solves the following problem:

$$k_t^k = \arg\max\left\{\Pi_t^k = \widehat{p}^k \widehat{H}_t^k : k = 1, 2, ..., K\right\} \tag{7}$$

This expression does not suggest that the fisherman captures in a single patch on each trip, but they can indeed capture in a single patch at a time, thus allowing them to choose several ones in a hierarchical order on each trip.

N ships are considered that capture a homogeneous species. The array of choice by each fisherman corresponds to the patches that are feasible spatially and politically; namely, considering that although some patches are geographically accessible, they may be forbidden from capturing activities.

### 2.3. Econometric model

The econometric estimation requires the decomposition of the model into two components: economic and biological. On the one hand, the economic component includes the capture function and the decision made by the fishermen about the patch in which their captures are performed. On the other hand, the biological model is composed by the equation that defines the migration of the stock between patches. The estimation of the model is carried out in two stages.

Both economic and biological components are linked through the biomass level existing in each patch. This is an explicative variable of the capture function in each patch and it also determines the benefits obtained by the fishermen in each fishing area. However, this variable is not observed in the data; instead, it is assumed that it is known by the fishermen at the time of making their decision. This allows estimating the biomass level that explains both captures and decisions on fishing areas by the fishermen.

Thus, in a first stage the capture function and the demand per patch are jointly estimated, also obtaining an estimated biomass level per patch and period. These biomass levels are used in a second estimation stage in order to obtain the parameters of both biological dispersion and growth model of the biomass.

In order to perform the estimation of the demand, the logit model of demand with unobserved characteristics proposed by Berry (1994) is used. In order to adapt the model to our estimation problem, a market is defined. Each of the home ports of the fishing vessels in each month. Each of the fishing patches is defined as product and as alternative option: not to perform captures in such month.

Based on the model expressed in the previous section, the linear random utility function that explains the decision of the fishermen is defined as:

$$u_{tk} = \ln \Pi_t^k + \beta_1 \ln\left(\widehat{p}_t^k\right) + \beta_2 \ln\left(z_t^k\right) + \beta_3 \ln\left(\widehat{x}_t^k\right) + \varepsilon_t^k \tag{8}$$

This equation can be interpreted in terms of the discrete selection models as:

$$u_{tk} = \alpha_0^I + \alpha_1^I \ln\left(p_t^k\right) + \alpha_2^I \ln\left(z_t^k\right) + \xi_t^k + \varepsilon_t^k \tag{9}$$

Where $\alpha_0^I$, $\alpha_1^I$ and $\alpha_2^I$ represents parameters of reference on the characteristics observed and the superscript $I$ is referred to the fact that they are parameters estimated in the first stage. $\xi_t^k$ represents characteristics not observed in the patches. These are assumed to be proportional to the logarithm of the biomass ($\beta \ln x_t^k$) which are thought independent from the $z$ vector; and $\varepsilon_t^k$ is the random error for the prediction of the capture by the fisherman. It is assumed that it has an Extreme Value distribution.

The non-observed features are estimated in the first step which, once identified, represent the estimated biomass for every fisherman. This biomass is the one used in the second step of the estimation.

According to the method pointed out by Berry (1994), the structural parameters of the above demand model can be estimated at the aggregate level by means of a function on the market participation of each patch in the characteristics observed for each market. In our case, this implies the estimation of a system of equations, one for each port of origin, where each observation corresponds to the monthly observations of both captures and fishing destinations. Thus, the equation to be estimated for each patch is as follows:

$$\ln(s_k) - \ln(s_0) = \alpha_0^I + \alpha_1^I \ln\left(p_t^k\right) + \alpha_2^I \ln\left(z_t^k\right) + \xi_t^k \qquad (10)$$

Where $s_k$ corresponds to the percentage of fishermen that decided to capture in the patch $k$ in the considered period, and $s_0$ is the aggregated percentage of fishermen that decided not to go fishing during that period. Following Berry *et al.* (1995), the left-side construction of the equation (10), estimated by means of a logit, will provide an estimate of the unknown parameters $\xi_t^k$.

In order to prove the method proposed, the first stage of our model considers a system of equations where five equations –such as the one in the previous case– are included; four equations for each patch, and other equation with the capture function. As in the previous case, there is one for each port of origin. These are jointly estimated with the capture function through the use of a non-linear SURE system (Zellner, 1962). Additionally it is possible to use information on the estimated biomass for each year in order to identify exactly the value of the parameter $\xi$ present in equation (9) and the biomass levels in each patch $\hat{x}_t^k$.

In the second stage, estimation coefficients of the previous biomass are used in order estimate the growth equation parameters and biomass migration presented in equations (1), (2) and (3). These can be rewritten for estimation purposes as:

$$\hat{x}_{t+1}^k = \alpha_0^{II} \hat{x}_t - \alpha_1^{II} \left(\hat{x}_t\right)^2 + \sum_{\substack{h=1 \\ h \neq k}}^{K} d_{hk} \hat{x}_t^k + \varepsilon_t^{II} \qquad (11)$$

Were $\alpha_0^{II} = (r - d_{kk})$ and $\alpha_1^{II} = \dfrac{r}{K_k}$. The identification of the structural parameters, $d_{kk}$ and $K_k$ requires an auxiliary estimation. For this, it is assumed that $r$ is equal for all patches and the sum of the carrying capacities of each patch must be equal than the carrying capacity of the whole system. Thus, the values $r$ and $K = \sum_k K_k$ are estimated by using the data of the entire fishery. This allows for the identification of all biological parameters of each patch.

In order to estimate the system of equations of (11) type, one for each patch, a SURE equations system is also used.

## 3. STUDY CASE AND DATA

The developed model is applied on a trial basis to the mackerel fishery in Central Southern Chile. This is one of the most important fisheries in both Chile and the world. This fishery comprises the areas located between the Valparaiso and Puerto Montt Regions. For estimation purposes, all industrial vessels whose port of origin is San Antonio, Talcahuano, Coronel and Valdivia that performed any capture operation between 2001 and 2004 were considered. The selection of this period corresponds to the integrity and availability of the database for the researchers.

Fisheries in this zone during 2004 recorded a landing of *circa* 1.5 million tons, out of which 79% corresponds to Chilean jack mackerel.

In this area the landings of mackerel carried out mainly in four ports. Since 2002 to 2004, the ports of Talcahuano and Valdivia, concentrated 86% of the captures. In 2004 the Central Southern industrial fleet performed 2,785 fishing trips. 95% of such trips returned to port with effective capture. In spatial terms, between January and February the activity was concentrated between the localities of Constitución (35°20' S 72°25' W) and Talcahuano (36°43'30" S-73°6'40" W), without exceeding 100-129 nautical miles from the coast. Between March and July the activity moved to the Isla Grande in Chiloé (41°51'43" S-73°49'52" W). In mid-June and July the oceanic operation began, following the resource beyond 600 nautical miles until September, in which the fleet continued operating in the northwestern area off Chilean coasts. Finally, between October and November the capture was reduced, being concentrated in the coastal areas around Mocha Island (38°23'06" S-73°52'00" W). In December, the captures experience an upturn with the return to coastal areas off Coquimbo Region coasts. This can be noted in Figure 1 that shows the spatial distribution of the captures during 2004.

Some authors have studied the migration behaviors of the Chilean jack mackerel from a biological standpoint. Arcos *et al.* (2001) point out that the migration patterns of this species consider a significant fraction of the population and as they occur on a regular basis, as the result of the alternation between two or more separated habitats. Thus, these authors have pointed out that the movements from one habitat to another determine the seasonal behavior of the fishery in the Central Southern area, with higher captures in the Winter season (April to August in Chile), when the mackerel is more available in coastal waters for fattening reasons. On the other hand, lower captures take place during spring and summer seasons, as a result of the migration process to oceanic waters for spawning (September to March). Finally, these authors

FIGURE 1

SPATIAL DISTRIBUTION OF THE MACKEREL CAPTURES IN 2004

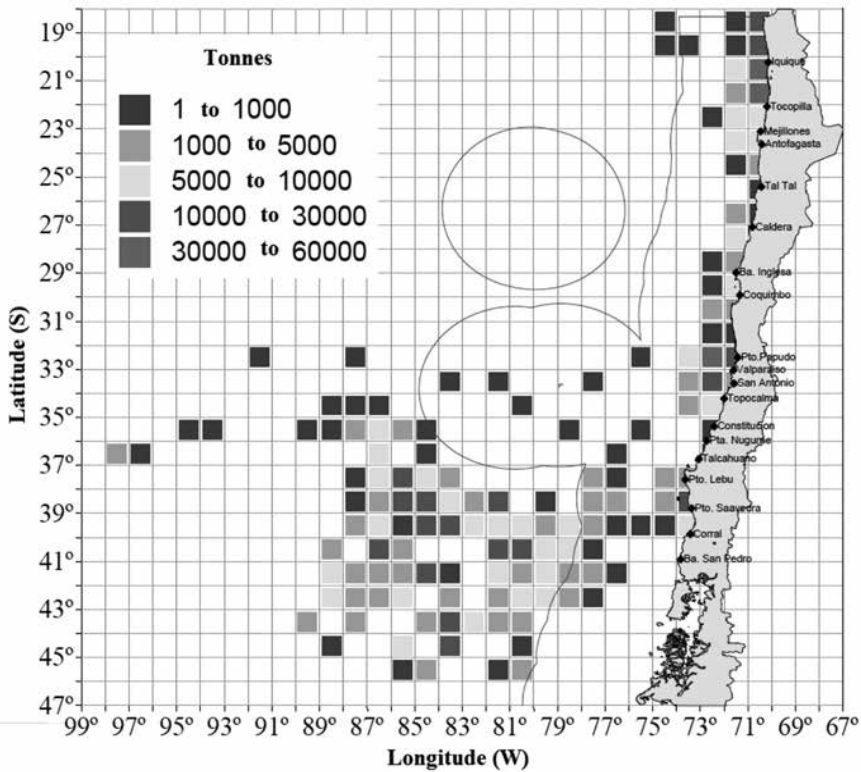

Source: SERNAPESCA.

acknowledge that other migration patterns may exist in different spatial and temporal scales. These patterns should be related to the use of the space as well as changes in the balance between benefits and costs of residence in each habitat.

Because of the migration characteristics of this species in the analyzed fishing zone, and the importance of this species for the Chilean fishery, this becomes an ideal area as a study case for the proposed model.

For both spatial and migration analysis the central southern area is divided into eight patches according to the classification used by the Chilean Undersecretary of Fishing (SERNAPESCA) to define the Chilean fishing areas. Figure 2 shows the location of the central southern zone of Chile as well as the eight patches taken into account for the analysis. The four ports of origin from which the extractive industrial activity is performed are also indicated.

FIGURE 2

CHILEAN MAP THAT CONSIDERS PATCHES AND PORTS DEFINED BY SERNAPESCA

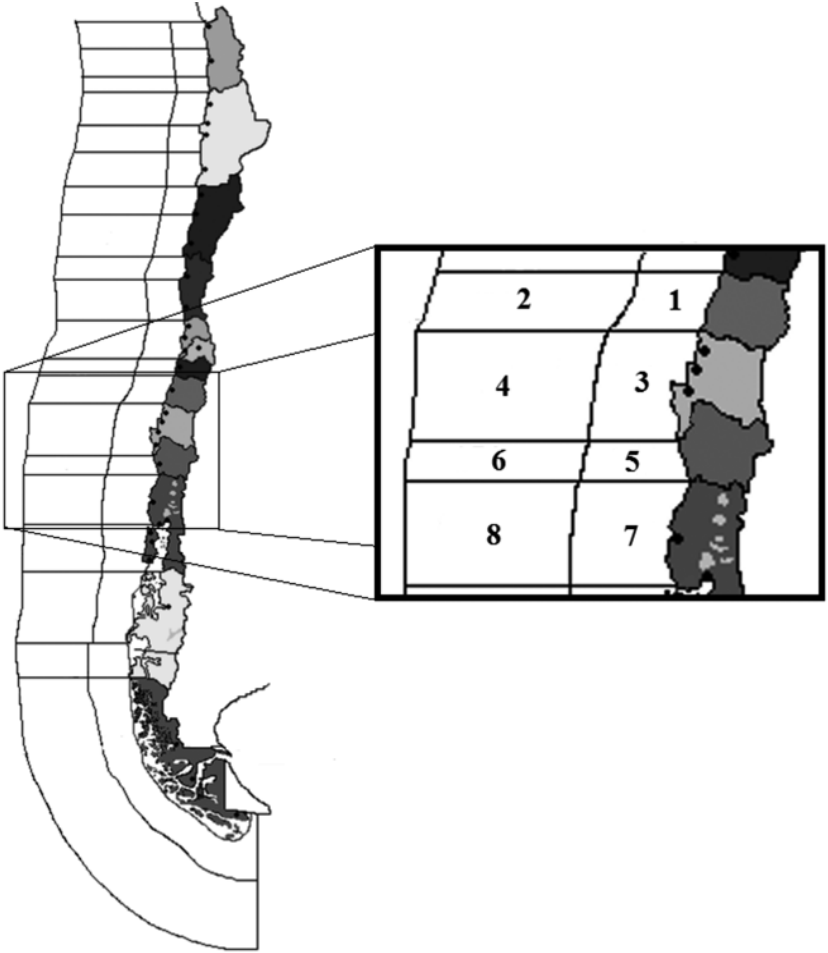

Note: Patches used in the analysis are highlighted.
Source: own.

The information comes from the landing records from the Chilean Undersecretary of Fishing, and includes the number of monthly trips and the landings per fishing zone for each of the vessels that operated in the fishery between 2001 and 2004. The database shows the variability in both average capture per patch and month. These can be observed in Figure 3 and Figure 4.

## FIGURE 3

AVERAGE CAPTURE PER PATCH AND YEAR (IN THOUSANDS OF TONS)

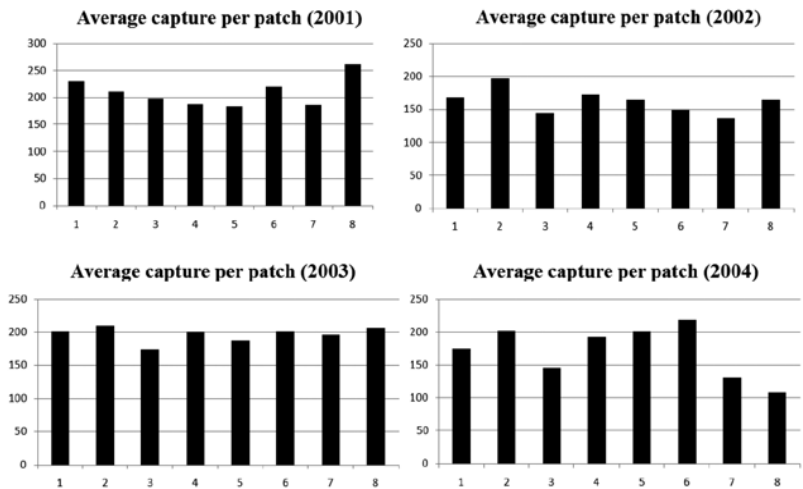

Source: SERNAPESCA.

## FIGURE 4

AVERAGE CAPTURE PER MONTH AND YEAR

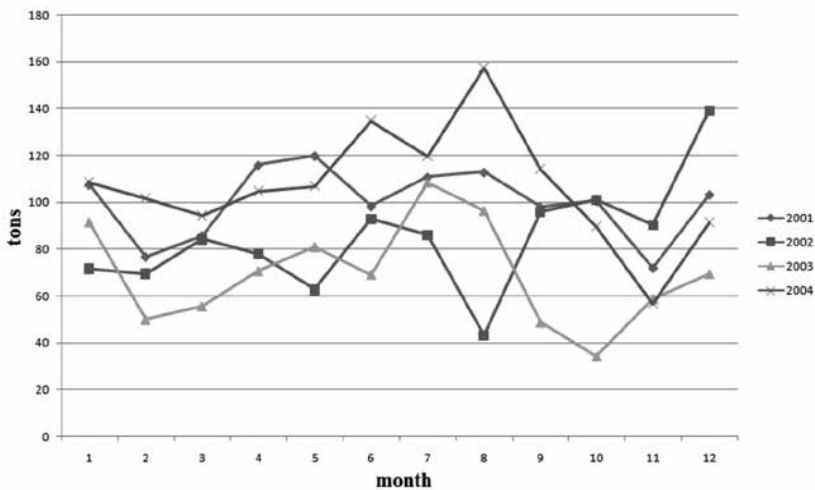

Source: SERNAPESCA.

The database possesses daily information from the Chilean Marine Fisheries Service (SERNAPESCA henceforth). It contains 35,063 observations with 113 vessels of diverse technical features. The data were aggregated monthly for the estimation (and for the practical use of dummy variables). Said vessels had *trachurus murphyi* as their sole target species.

TABLE 1

DESCRIPTIVES FOR TRAVEL PER YEAR

|  | 2001 | 2002 | 2003 | 2004 |
|---|---|---|---|---|
| Total travel | 27 774 | 36 730 | 34 947 | 32 560 |
| Mean travel per vessel | 100.408 | 83.98 | 69.20 | 104.45 |
| SD travel per vessel | 14.52 | 22.60 | 20.86 | 23.82 |

Source: SERNAPESCA.

## 4. RESULTS

The results in the reduced form of the parameters of the estimation of the discrete selection model in step 1 are presented in Table 2. The parameters $\alpha_{0,t}^{I}$ show the coefficient corresponding to the constant for each year $t$. Parameters with the form $\alpha_{1,t,j}^{I}$ correspond to the coefficients associated to the net benefits of capture per ton in each patch, in year $t$, for a ship that departs from port $j$. Given the fact that the distances to the patches are different depending from the port of origin, it is considered that this parameter is also different for each port. The parameters $\alpha_{2,t}^{I}$ correspond to the exponent of the fishing aggregated effort on the function of captures in each year. In this stage, biomasses per patch and month are estimated.

The net benefit corresponds to the difference between the price and cost per capture, all measured in tons. Consequently, the parameters $\alpha_{1,t}^{I}$ are interpreted as the net benefits for going fishing from any of the four ports in each year.

By using dummy variables per patch and year, it is possible to identify the biomass as $d_t^k = \beta \ln(x_t^k)$. The estimated biomasses are presented in Figure 5.

With these coefficients estimated in the second stage we proceed to estimate the biological migration model.

In the Table 3 we present the results for the estimation of reduced form of the biological migration model.

The parameters of the form $\alpha_{0,k}^{II}$ match the coefficients corresponding to the variable $x_{tk}$ used to identify $r$ and $d_{kk}$. In turn, the parameters of form $\alpha_{1,k}^{II}$ match the coefficients corresponding to the variable $x_{t,k}^2$ for each of the eight patches ($k$) considered in the study, and it's required to identify the charge capacity.

Parameters in reduced form obtained from the first and second steps of estimations allow generating the structural parameters of both economic and biological models; particularly those parameters associated to both capture and migration functions

TABLE 2

RESULTS FROM THE ESTIMATION REDUCED FORM, STEP 1

|  | Parameter | St. Error | z-value |
|---|---|---|---|
| $\alpha^I_{0,2001}$ | 4.8311 | 0.0576 | 83.86 |
| $\alpha^I_{0,2002}$ | 5.0050 | 0.0579 | 86.43 |
| $\alpha^I_{0,2003}$ | 5.0164 | 0.0474 | 105.86 |
| $\alpha^I_{0,2004}$ | 4.7792 | 0.0492 | 97.11 |
| $\alpha^I_{1,2001,1}$ | 0.1500 | 0.0138 | 10.88 |
| $\alpha^I_{1,2001,2}$ | 0.0281 | 0.0037 | 7.62 |
| $\alpha^I_{1,2001,3}$ | 0.0612 | 0.0036 | 17.09 |
| $\alpha^I_{1,2001,4}$ | 0.0059 | 0.0012 | 4.99 |
| $\alpha^I_{1,2002,1}$ | 0.1510 | 0.0138 | 10.98 |
| $\alpha^I_{1,2002,2}$ | 0.0292 | 0.0036 | 8.01 |
| $\alpha^I_{1,2002,3}$ | 0.0614 | 0.0036 | 16.90 |
| $\alpha^I_{1,2002,4}$ | 0.0057 | 0.0012 | 4.81 |
| $\alpha^I_{1,2003,1}$ | 0.1450 | 0.0138 | 10.51 |
| $\alpha^I_{1,2003,2}$ | 0.0297 | 0.0036 | 8.25 |
| $\alpha^I_{1,2003,3}$ | 0.0636 | 0.0036 | 17.92 |
| $\alpha^I_{1,2003,4}$ | 0.0061 | 0.0011 | 5.36 |
| $\alpha^I_{1,2004,1}$ | 0.1460 | 0.0137 | 10.65 |
| $\alpha^I_{1,2004,2}$ | 0.0268 | 0.0036 | 7.55 |
| $\alpha^I_{1,2004,3}$ | 0.0596 | 0.0035 | 17.16 |
| $\alpha^I_{1,2004,4}$ | 0.0056 | 0.0011 | 4.99 |
| $\alpha^I_{2,2001}$ | 1.0798 | 0.0133 | 80.96 |
| $\alpha^I_{2,2002}$ | 1.0464 | 0.0132 | 79.30 |
| $\alpha^I_{2,2003}$ | 1.0538 | 0.0109 | 96.44 |
| $\alpha^I_{2,2004}$ | 1.0940 | 0.0115 | 95.10 |
|  | Obs | Parameters | $R^2$ |
| Equation 1 | 906 | 98 | 0.79 |
| Equation 2 | 906 | 102 | 0.08 |
| Equation 3 | 906 | 102 | 0.45 |
| Equation 4 | 906 | 102 | 0.24 |
| Equation 5 | 906 | 102 | 0.75 |

Source: Own.

between patches. Thus, Table 4 shows the estimated parameters of the capture function and Table 5 shows the estimations of migration and growth parameters.

Results in Table 4 show the parameters associated to the technological constant of the capture function, as well as the elasticity of the aggregated fishing effort. It can be observed that the elasticity is slightly higher than one and significantly higher than this value. All economic variables are statistically significant and they showcase the expected signs and magnitudes according to the literature; e.g., on the one hand, the capture is positive and its magnitude is expressed in tons. On the other hand, the elasticity shows that if the net benefits per patch increases, then the amount demanded for that patch is increased as well.

FIGURE 5

ESTIMATED BIOMASS PER MONTH AND PATCH

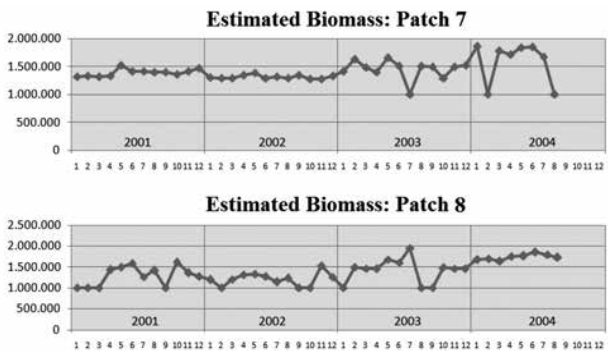

Source: own.

## TABLE 3

RESULTS FROM THE ESTIMATION REDUCED FORM, STEP 2

|  | Parameter | St. Error | z-value |
|---|---|---|---|
| $\alpha^{II}_{0,1}$ | -0.03547 | 0.15802 | -0.22 |
| $\alpha^{II}_{0,2}$ | -0.07644 | 0.19494 | -0.39 |
| $\alpha^{II}_{0,3}$ | 0.86710 | 0.09543 | 9.09 |
| $\alpha^{II}_{0,4}$ | 0.86040 | 0.14392 | 5.98 |
| $\alpha^{II}_{0,5}$ | -0.11972 | 0.18741 | -0.64 |
| $\alpha^{II}_{0,6}$ | 0.07991 | 0.12954 | 0.62 |
| $\alpha^{II}_{0,7}$ | 0.62316 | 0.13785 | 4.52 |
| $\alpha^{II}_{0,8}$ | -0.83693 | 0.24211 | -3.46 |
| $\alpha^{II}_{1,1}$ | -9.87E-08 | 5.56E-08 | -1.77 |
| $\alpha^{II}_{1,2}$ | -4.51E-07 | 7.30E-08 | -6.18 |
| $\alpha^{II}_{1,3}$ | -5.68E-07 | 4.90E-08 | -11.58 |
| $\alpha^{II}_{1,4}$ | -7.26E-07 | 4.65E-08 | -15.62 |
| $\alpha^{II}_{1,5}$ | -1.93E-07 | 6.34E-08 | -3.05 |
| $\alpha^{II}_{1,6}$ | -3.49E-07 | 4.98E-08 | -7.00 |
| $\alpha^{II}_{1,7}$ | -5.46E-07 | 6.50E-08 | -8.39 |
| $\alpha^{II}_{1,8}$ | 2.34E-08 | 9.42E-08 | 0.25 |
|  | Obs | Parameters | $R^2$ |
| Equation 1 | 571 | 4 | 0.38 |
| Equation 2 | 571 | 4 | 0.62 |
| Equation 3 | 571 | 5 | 0.28 |
| Equation 4 | 571 | 6 | 0.34 |
| Equation 5 | 571 | 5 | 0.24 |
| Equation 6 | 571 | 6 | 0.44 |
| Equation 7 | 571 | 5 | 0.16 |
| Equation 8 | 571 | 6 | 0.20 |

Source: own.

TABLE 4

ESTIMATION OF PARAMETERS CAPTURE FUNCTION

|  | Parameter | St. Error | z-value |
|---|---|---|---|
| $\gamma_{2001}$ | 125.352 | 1.0593 | 83.86 |
| $\gamma_{2002}$ | 149.160 | 1.0596 | 86.43 |
| $\gamma_{2003}$ | 150.864 | 1.0485 | 105.86 |
| $\gamma_{2004}$ | 119.006 | 1.0504 | 97.11 |
| $\alpha_{2001}$ | 1.07982 | 0.0133 | 80.96 |
| $\alpha_{2002}$ | 1.04644 | 0.0132 | 79.30 |
| $\alpha_{2003}$ | 1.05384 | 0.0109 | 96.44 |
| $\alpha_{2004}$ | 1.09403 | 0.0115 | 95.10 |

Source: own.

TABLE 5

ESTIMATED BIOLOGICAL PARAMETERS OF GROWTH AND MIGRATION

|  | Parameter | St. Error | t-value |
|---|---|---|---|
| r | 0.02868 | 0.0548 | 5.24 |
| $d_{11}$ | 0.32227 | 0.1580 | -0.22 |
| $d_{22}$ | 0.36324 | 0.1949 | -0.39 |
| $d_{33}$ | -0.58030 | 0.0954 | 9.09 |
| $d_{44}$ | -0.57359 | 0.1439 | 5.98 |
| $d_{55}$ | 0.40652 | 0.1874 | -0.64 |
| $d_{66}$ | 0.20689 | 0.1295 | 0.62 |
| $d_{77}$ | -0.33636 | 0.1378 | 4.52 |
| $d_{88}$ | 1.12373 | 0.2421 | -3.46 |
| $d_{12}$ | -0.03225 | 0.0230 | -1.40 |
| $d_{13}$ | 0.17289 | 0.0764 | 2.26 |
| $d_{21}$ | 0.56443 | 0.0803 | 7.02 |
| $d_{24}$ | 0.02172 | 0.0805 | 0.27 |
| $d_{31}$ | -0.10329 | 0.0297 | -3.47 |
| $d_{34}$ | 0.06132 | 0.0304 | 2.02 |
| $d_{35}$ | -0.03265 | 0.0436 | -0.75 |
| $d_{42}$ | -0.28887 | 0.0252 | -11.48 |
| $d_{43}$ | 0.43961 | 0.0758 | 5.80 |
| $d_{46}$ | -0.07317 | 0.0249 | -2.94 |
| $d_{53}$ | 0.43810 | 0.0778 | 5.63 |
| $d_{56}$ | -0.10528 | 0.0296 | -3.56 |
| $d_{57}$ | 0.03319 | 0.0877 | 0.38 |
| $d_{64}$ | 0.20394 | 0.0531 | 3.84 |
| $d_{65}$ | 0.29074 | 0.0538 | 5.40 |
| $d_{68}$ | -0.15976 | 0.0548 | -2.92 |
| $d_{75}$ | 0.16404 | 0.0494 | 3.32 |
| $d_{78}$ | -0.03292 | 0.0282 | -1.17 |
| $d_{86}$ | 0.87619 | 0.1177 | 7.45 |
| $d_{87}$ | -0.10154 | 0.0463 | -2.19 |

Source: own.

The Table 5 shows the parameters associated to the migration of the stock, identified from the second stage of estimation. Results indicate that an important fraction of the migration parameters are statistically significant. *A priori*, no restriction has been imposed to the signs of these parameters, allowing the data to indicate unrestricted migration parameters. Results associated to these migration patterns are presented in Figure 6.

Regarding the identification of the carrying capacity per patch, unfortunately one of the parameters in reduced form associated to the carrying capacity of the

FIGURE 6

MIGRATION PARAMETERS ESTIMATED IN A REPRESENTATION OF PATCHES

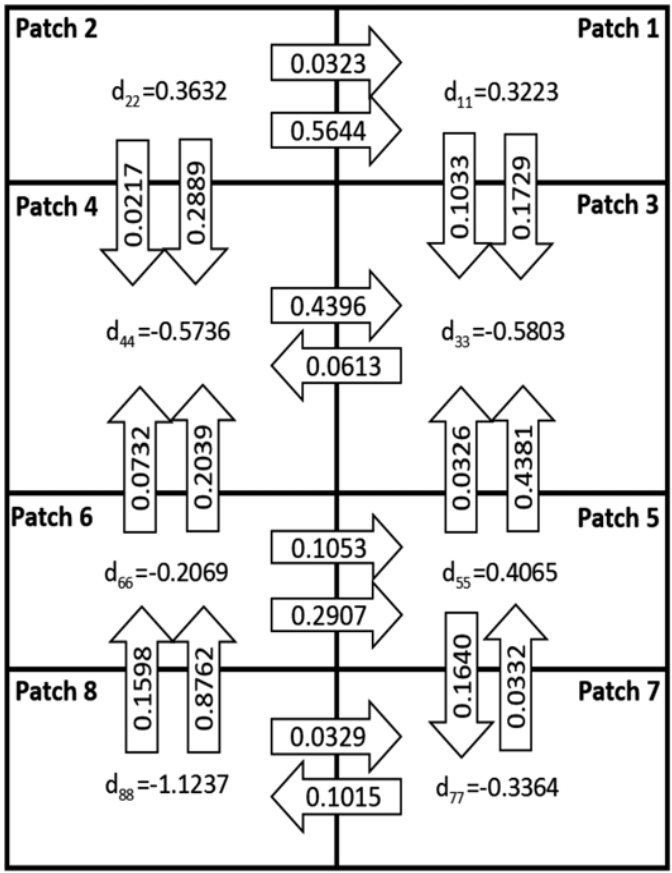

Source: own.

TABLE 6

CARRYING CAPACITY PER EACH PATCH

| Patch | Upper limit confidence interval | | Mean |
|---|---|---|---|
| | At 80% | At 90% | Estimated |
| $K_1$ | 6.736 | 6.987 | 19.045 |
| $K_2$ | 2.105 | 2.325 | 4.168 |
| $K_3$ | 1.815 | 2.045 | 3.309 |
| $K_4$ | 1.459 | 1.655 | 2.589 |
| $K_5$ | 4.179 | 4.470 | 9.740 |
| $K_6$ | 2.779 | 3.080 | 5.386 |
| $K_7$ | 1.821 | 2.033 | 3.443 |
| $K_8$ | 11.757 | 10.056 | -80.330 |

Source: own.

system has a negative sign, which is statistically significant and prevents the correct identification of the carrying capacities of all patches. This is due to the fact that the sum of the carrying capacities must equate the total carrying capacity of the system. This is shown in Table 6, which indicates that the procedure discussed in the previous section presents a negative carrying capacity for the entire system, if the estimated average value of the parameters is assessed.

In order to solve the above problem the estimation of the carrying capacities is taken into account, by using the upper limits of the confidence intervals of the estimated parameters because they are positive, thus meeting the estimation requirements of the carrying capacities for each patch. Results are presented in Table 6. It is observed that this allows obtaining the estimations on the carrying capacity sensitive to the limit used.

## 5. CONCLUSIONS

In this study, it has been proposed and applied a method for the estimation of a bioeconomic model that considers the migration of the stock between patches. The model has been estimated for the jack mackerel fishery in central southern Chile.

By using a two SUR estimation models, it is possible to identify migration parameters across eight patches, in four years, from four ports, thanks to the use of a simultaneous method which will allow incorporating all this information.

In the first step, a capture function is used, with one equation for the demand of each patch. One regression per each year will allow obtaining the biomass estimated per patch. In the base example, the biomass estimated is consistent with the observations made by the National System of Statistical Fishery of Chile.

In the second step, another SUR model is deployed by using a dispersion matrix and by identifying the carriage capacity. Once the database is set, the behavioral pattern of the biomass can be detected; e.g., from patch eight to patch four, and from patch two to patch four. In addition, it is perceived that the biomass is close to the coast in patches one and five. Finally, in patches three and seven the biomass migrates offshore.

In this particular case, the results indicate that with the model presented in this work it is possible to calculate with statistical significance the parameters in function of captures as well as the migration parameters between patches. However, it is not possible to clearly identify the carrying capacities of each patch, at least for the sample available. Of course, the results obtained can be used to perform analyses of the spatial behavior of the fishing fleet as well as biological and economic models on policies of spatial management. By using a model demand with non-observed characteristics as proposed by Berry *et al.* (1998) with aggregated data such as per-patch capture; and by knowing the origin port, the spatial behavior can be obtained.

The model can be also tested in other fisheries, especially those with broad patterns of migration. All these possible analyses, although they were outside the scope of this investigation, constitute an opportunity to continue deepening on this sort of analyses in fisheries in further investigations.

Possibly, the weakness in this proposal may lie in the necessity for a richer database to be estimated. In fact, the serial time must be continual and balanced in each patch. In other cases, this methodology can be applied to other sea species with migratory behaviors more or less similar to the ones analyzed in this work.